# Magnetic Field Effect on Electron Spin Dynamics in (110) GaAs Quantum wells

G. Wang[1,2], A. Balocchi[2], A. V. Poshakinskiy[3], C. R. Zhu[1], S. A. Tarasenko[3], T. Amand[2], B. L. Liu[1] and X. Marie[2,4]

[1]Beijing National Laboratory for Condensed Matter Physics, Institute of Physics, Chinese Academy of Sciences, P.O. Box 603, Beijing 100190, China
[2]Université de Toulouse, INSA-CNRS-UPS, LPCNO, 135 avenue de Rangueil, 31077 Toulouse, France
[3]A.F. Ioffe Physical-Technical Institute, 194021 St. Petersburg, Russia
E-mail: marie@insa-toulouse.fr

**Abstract.** We study the electron spin relaxation in both symmetric and asymmetric GaAs/AlGaAs quantum wells (QWs) grown on (110) substrates in an external magnetic field $B$ applied along the QW normal. The spin polarization is induced by circularly polarized light and detected by time-resolved Kerr rotation technique. In the asymmetric structure, where a δ-doped layer on one side of the QW produces the Rashba contribution to the conduction-band spin-orbit splitting, the lifetime of electron spins aligned along the growth axis exhibits an anomalous dependence on $B$ in the range $0<B<0.5$ T; this results from the interplay between the Dresselhaus and Rashba effective fields which are perpendicular to each other. For larger magnetic fields, the spin lifetime increases, which is the consequence of the cyclotron motion of the electrons and is also observed in (001)-grown quantum wells. The experimental results are in agreement with the calculation of the spin lifetimes in (110)-grown asymmetric quantum wells described by the point group Cs where the growth direction is not the principal axis of the spin-relaxation-rate tensor.

## 1. Introduction

The electron spin relaxation in semiconductor quantum wells with no inversion symmetry is dominated by the D'yakonov-Perel' mechanism [1]. It results from the non-degenerate conduction band (CB) electron spin states originating from the combined effect of Spin-Orbit (SO) interaction and inversion asymmetry; the corresponding energy splitting can be viewed as a consequence of the action of an effective magnetic field whose amplitude and orientation varies with the electron wave vector $\boldsymbol{k}$. The electron spin thus experiences a fluctuating effective magnetic field when the electron wave vector changes yielding a relaxation time $\tau_s$ for a given spin component

$$\tau_s^{-1} \sim \left\langle \Omega_{k,\perp}^2 \right\rangle \tau_p^*, \tag{1}$$

where $\left\langle \Omega_{k,\perp}^2 \right\rangle$ is the mean square component of the precession vector $\boldsymbol{\Omega_k}$ corresponding to the CB spin-splitting in the plane orthogonal to the considered spin direction and $\tau_p^*$ the momentum relaxation time for an individual electron.

---

[4] Author to whom any correspondence should be addressed.

The CB spin splitting is mainly due to two additive contributions: the Bulk Inversion Asymmetry (BIA or Dresselhaus contribution [2]) and the Structural Inversion Asymmetry (SIA or Rashba contribution [3]). The latter can be induced by an external voltage applied to the heterostructure. The Dresselhaus contribution depends drastically on the spatial symmetry of the structure [4]. As a consequence, the electron spin relaxation tensor varies strongly if the quantum wells are grown on (001), (110), or (111) oriented substrates. The interplay between the BIA and SIA terms has been studied experimentally and theoretically by several groups in recent years [5-14].

For (110) quantum wells, the precession vector due to the BIA term is oriented along $z \parallel$ [110] direction and has the form [15]

$$\mathbf{\Omega}_{BIA}(\mathbf{k}) = \frac{\gamma}{\hbar} \langle k_z^2 \rangle (0, 0, k_x), \qquad (2)$$

where $\gamma$ is the Dresselhaus coefficient, $\langle k_z^2 \rangle$ is the averaged squared wave vector along the growth direction, and $x \parallel [1,\bar{1},0]$ and $y \parallel [0,0,\bar{1}]$ are the in-plane axes.

As a consequence, if the electron spin is also aligned along the growth direction (*i.e.* perpendicular to the QW plane), the DP spin relaxation mechanism is suppressed (in the absence of SIA). This theoretical prediction has been confirmed by different experimental groups which have probed the electron spin dynamics with time-resolved Kerr rotation or photoluminescence experiments [16-20]; electron spin relaxation times as long as 20 ns have been measured for electrons spins parallel to the [110] GaAs growth direction at T=300 K. Recent spin grating experiments showed the possibility of transporting electron spins in these structures over ~4 µm at room temperature [21] and operation of a Spin VCSEL based on (110) QWs has also been demonstrated [22].

If an electric field is applied parallel to the quantum well growth direction or an asymmetric doping generating a built-in electric field is present in the structure, the additional Rashba contribution to the spin splitting $\mathbf{\Omega}_{SIA}$ lying in the QW plane emerges [4]:

$$\mathbf{\Omega}_{SIA}(\mathbf{k}) = \frac{2r_{41}E}{\hbar} (k_y, -k_x, 0), \qquad (3)$$

where $\alpha = r_{41} E$ is the Rashba coefficient, $E$ is the electric field, and $r_{41}$ is a coefficient determined by QW design and chemical composition.

The total spin precession vector $\mathbf{\Omega}_k = \mathbf{\Omega}_{BIA} + \mathbf{\Omega}_{SIA}$ is thus tilted with respect to the QW growth axis and an electron spin oriented along the growth axis will therefore experience an efficient DP spin relaxation process. A clear tuning of the electron spin relaxation time with the applied electric field was demonstrated by Karimov *et al.* [10]. The electron spin relaxation time $\tau_s$ decreases from 1000 ps to 100 ps when the electric field varies from 20 to 80 KV/cm. The electric field dependence of the SIA term has allowed Eldridge *et al.* to extract the Rashba coefficient [23].

In optical orientation experiments, the application of a longitudinal external magnetic field (*i.e.* parallel to the propagation of light) usually yields an increase of the electron spin relaxation time, as described by many theoretical works [24-28] and observed in experiments on (001)-oriented structures [29,30]. The dominant mechanism of such a spin relaxation suppression is the cyclotron motion of electrons, which causes the direction of $k$ to change in a way equivalent to an increase in the relaxation time $\tau_p^*$. Additionally, the Larmor precession of electron spins around a strong enough external magnetic field should suppress the precession around the perpendicular internal random fields, thus preserving the initial orientation of electron spins (whatever the origin of this internal random field is). In the unique situation of (110) quantum wells, where the effective magnetic field $\Omega_{BIA}$ is also oriented parallel to the growth direction, the effect of a longitudinal magnetic field on the electron spin dynamics can be strongly different [31]. This effect however has not been studied experimentally so far.

In this paper we present an experimental and theoretical investigation of the magnetic field dependence of the electron spin relaxation time performed in symmetric and asymmetric (110) GaAs/AlGaAs quantum wells. The experimental results on the asymmetric quantum well reveal an unusual dependence on the magnetic field. For weak external magnetic fields, the decay of spin density after circularly polarized pump pulse is not mono-exponential and exhibits slow oscillations in time due to the interplay between the BIA and SIA terms and the external magnetic field. We obtain very good fits of the magnetic field dependences of the electron spin lifetimes using the Dresselhaus and Rashba parameters for GaAs.

In the following section we present the investigated samples and the experimental set-up which has been used to measure the magnetic field dependence of the electron spin dynamics. The experimental results on both symmetric and asymmetric QWs are summarized in section 3. The measured dynamics in (110)-oriented QW is compared to the one obtained on a similar QW grown on (001) substrate. Finally we present in section 4 the model which allows us to calculate the spin relaxation tensor in both symmetric and asymmetric QW and compare the calculations to the experimental results.

## 2. Samples and experimental Set-up

Three samples have been investigated in this work. Two structures were grown by molecular beam epitaxy on semi-insulating (110) substrates. The first sample (labeled Asym) contains a single GaAs/Al$_{0.3}$Ga$_{0.7}$As well with a width $L_W$ = 8 nm; a n-type $\delta$-doped layer has been inserted in the Al$_{0.3}$Ga$_{0.7}$As barrier at a distance $d$ =15 nm from the QW interface on the substrate side. This $\delta$-doped layer generates an asymmetric potential profile in the QW (inset of figure 1) yielding a SIA term in the conduction band [4], the n-type doping is ~ $8 \times 10^{11}$ cm$^{-2}$ [11,32]. The second sample (labeled Sym) has the same quantum well characteristics but with no $\delta$-doped layer ("rectangular" quantum well). For comparison, we have also studied the spin dynamics in a (001)-oriented rectangular quantum well structure with the same characteristics as the ones of the (110) samples: $L_W$ = 8 nm and 30% aluminum fraction in the barrier.

The electron spin dynamics in these three samples have been studied by Time Resolved Kerr Rotation (TRKR) experiments [8]. The samples are excited near normal incidence with degenerate pump and delayed probe pulses by a Ti:sapphire laser (pulse width: 100 fs and repetition frequency: 76 MHz). The circular polarization of the pump beam is modulated at a frequency of 50 kHz by a photo-elastic modulator. The Kerr rotation of the linearly polarized probe beam is detected as a function of the delay between the pump and the probe pulses. The pump and probe beams have average power of 10 and 0.5 mW respectively. The investigated samples are held in an Oxford magneto-optical cryostat supplied with a 7 T split-coil superconducting magnet and the experiments are performed in the Faraday configuration: the magnetic field is applied along the growth direction of the quantum wells.

The Kerr rotation signal, proportional to the electron spin density [33], decays with a characteristic time $T_s$ which writes: $1/T_s = 1/\tau_s + 1/\tau_r$, where $\tau_s$ is the electron spin relaxation time and $\tau_r$ is the carrier lifetime (the very short hole spin relaxation time can be neglected. The experimental results presented in this paper have been obtained at T=150 K, a temperature high enough to avoid excitonic effects which could complicate the interpretation of the data. We have measured the carrier lifetime by time-resolved photoluminescence experiments; we found $\tau_r \sim 2000$ ps at T=150 K (not shown) [19,16,17]. We have checked that the intrinsic birefringence of (110) quantum wells does not affect our experimental results [34]. It was shown recently that a spin-independent Kerr rotation due to nonlinear pump-induced birefringence can occur in time-resolved Kerr rotation experiments [35]. This effect is cancelled in our experimental configuration thanks to the 50 kHz modulation of the pump beam: the Kerr rotation dynamics displayed in figure 1 are independent of the orientation of the linearly polarized probe beam with respect to the crystal orientation (not shown).

## 3. Experimental results

Figure 1 presents the normalized Kerr Rotation Dynamics for the symmetric and asymmetric quantum well structures. As expected, we observe a much faster decay of the TRKR signal in the asymmetric quantum well ($T_s \sim 170$ ps) compared to the symmetrical one ($T_s \sim 1100$ ps). In this optical orientation experiment the photogenerated electron spin is initially aligned along the growth direction of the quantum well. In the symmetric QW it is thus parallel to $\mathbf{\Omega}_{BIA}$ yielding a suppression of the DP spin relaxation and therefore a long spin relaxation time is observed in agreement with previous measurements. Because of the presence of the electric field along the growth direction induced by the δ-doping, the electron spins in the asymmetric QW experience in contrast a tilted effective magnetic field resulting from the combined action of the Dresselhaus and Rashba effects: $\mathbf{\Omega}_k = \mathbf{\Omega}_{BIA} + \mathbf{\Omega}_{SIA}$. The DP mechanism is here efficient for the $S_z$ spin component leading to the much shorter measured spin relaxation time.

The excitation power dependence of the spin lifetime in the asymmetric QW displayed in figure 2a confirms that it is governed by the DP mechanism. We observe an *increase* of $T_s$ from ~170 to 200 ps when the excitation power increases from 5 to 50 mW (note that this spin density lifetime $T_s$ measured by TRKR corresponds here to the electron spin relaxation time $\tau_s$ since $\tau_s \ll \tau_r$. This increase corresponds to the motional narrowing effect induced by the shorter momentum relaxation time occurring at higher photogenerated electron densities. In the symmetric QW such an increase is not observed which is consistent with the fact that the DP mechanism was suppressed since the electron spins are parallel to the Dresselhaus effective magnetic field $\mathbf{\Omega}_{BIA}$. On the contrary we observe a *decrease* of $T_s$ when the excitation power increases as shown in figure 2b. This is in agreement with previous reports which attribute this decrease to an increased efficiency of the Bir-Aronov-Pikus mechanism (exchange interaction) induced by the larger hole photogenerated density [16,35,20,36].

Figure 3a presents the measured electron spin dynamics in the asymmetric quantum well for different longitudinal magnetic field amplitudes (from $B = 0$ to 1 Tesla). The main effect is an increase of the spin relaxation time with $B$. A careful analysis shows that (*i*) the spin dynamics cannot be characterized by a simple mono-exponential decay time whatever the magnetic field is and (*ii*) a crossing point is evidenced around $t \sim 400$ ps, see the inset of figure 3a (*i.e.* for $B = 0.3$ T, the electron spin polarization is smaller for $t > 400$ ps than the one for $B = 0$ whereas the opposite behavior occurs for $t < 400$ ps). Though these effects are not large, their reproducibility has been checked by performing several experimental runs. We show in the next section that such a behavior follows exactly the predictions from the microscopic theory.

For comparison, we present in figure 4a the variation of $T_s$ as a function of $B$ in a symmetric (001) GaAs/AlGaAs QW having the same aluminium fraction and same well width $L_W$= 8 nm. In this case, $\Omega_{BIA}$ lies in the QW plane and we clearly observe a monotonous increase of the electron spin relaxation time with $B$ as a consequence of the cyclotron effect[30,37]. We emphasize that the spin relaxation time is here strictly mono-exponential whatever the magnetic field is in the range $B$ = 0-1 T. This is a key difference compared to the asymmetric quantum well behavior where a slight oscillation can be observed for the curves obtained in the range $0.1 < B < 0.4$ T (figure 3a).

## 4. Model and discussions

In general, the external magnetic field has two effects on the electron spin dynamics. First, it results in the Larmor precession of the electron spin around the magnetic field. Second, the magnetic field leads to cyclotron motion of electrons, which causes the direction of $k$ to dynamically change thereby suppressing the DP mechanism and modifying the Larmor frequency[25].

For the collision-dominated regime of spin dynamics, relevant to our experiments at T=150 K, both effects can be treated quasi-classically in the framework of the spin-density-matrix technique. The final equation describing the time evolution of the optically injected electron spin has the form [31]

$$\frac{dS_\alpha}{dt} + [\mathbf{S} \times \mathbf{\Omega}'_L]_\alpha = G_\alpha - \sum_\beta \Gamma_{\alpha\beta}(B_z) S_\beta - \frac{S_\alpha}{\tau_r}, \quad (4)$$

where $\mathbf{\Omega}'_L = \mathbf{\Omega}_L + \delta\mathbf{\Omega}_L$ is the Larmor frequency modified by cyclotron motion, $\Omega_{L,\alpha} = g_{\alpha\beta} \frac{\mu_0 B_\beta}{\hbar}$, $g_{\alpha\beta}$ is the g-factor tensor in the absence of cyclotron motion, $\mu_0$ is the Bohr magneton,

$$\delta\mathbf{\Omega}_L = -\frac{1}{2}\left\langle \frac{\omega_c \tau_p^*}{1+(\omega_c \tau_p^*)^2} \mathbf{\Omega}_k \times \frac{\partial \mathbf{\Omega}_k}{\partial \varphi_k} \right\rangle, \quad (5)$$

$\varphi_k$ is the polar angle of the wave vector $k$, $\omega_c = eB_z/(m^*c)$ is the cyclotron frequency, $m^*$ is the effective mass, $\Gamma_{\alpha\beta}(B_z)$ is the tensor of spin dephasing rates slowed-down by cyclotron motion,

$$\Gamma_{\alpha\beta}(B_z) = \left\langle \frac{\tau_p^*(\Omega_k^2 \delta_{\alpha\beta} - \Omega_{k,\alpha}\Omega_{k,\beta})}{1+(\omega_c \tau_p^*)^2} \right\rangle, \quad (6)$$

$G$ is the spin generation rate due to optical orientation, and the angular brackets in Eqs. (5) and (6) denote the averaging $\langle A \rangle = \int A\left(\frac{df_\varepsilon}{d\varepsilon}\right)d\mathbf{k} / \int \left(\frac{df_\varepsilon}{d\varepsilon}\right) d\mathbf{k}$ with $f_\varepsilon$ being the electron distribution function.

In (001)-grown quantum wells, the spin-orbit effective magnetic field lies in the QW plane, see Fig. 6a. The precession vector due to the BIA term, which is relevant for symmetric QWs, has the form [1]:

$$\Omega_{BIA}(\mathbf{k}) = \frac{2\gamma}{\hbar} \langle k_z^2 \rangle (k_y, k_x, 0), \tag{7}$$

where we use the coordinate frame $x \parallel [1,\bar{1},0]$, $y \parallel [1,1,0]$, and $z \parallel [0,0,1]$ for (001)-oriented QWs. The growth axis $z$ is one of the principal axes of the spin relaxation rate tensor and $g$-factor tensor. Therefore, the effect of the longitudinal magnetic field on the spin relaxation governed by D'yakonov-Perel' mechanism simply consists in the slowdown of spin dephasing due to cyclotron motion. The dependence of the spin relaxation time for spins aligned along the growth direction $\tau_z(B_z) = 1/\Gamma_{zz}(B_z)$ on magnetic field has the form

$$\tau_z(B_z) = \frac{\tau_z(0)}{1 + \omega_c^2 \tau_p^{*2}}, \tag{8}$$

where $\tau_z(0)$ is the spin relaxation time in zero magnetic field. For the Boltzmann distribution of electrons with the temperature $T$, $\tau_z(0)$ is given by

$$\frac{1}{\tau_z(0)} = \frac{8 m^* \gamma^2 <k_z^2>^2 \tau_p^*}{\hbar^4} k_B T. \tag{9}$$

For $L_W = 8$ nm, the calculation with finite barrier height yields $\langle k_z^2 \rangle = 4.3 \times 10^{-4}$ Å$^{-2}$. Using $\gamma =$ eV· Å$^3$ as recently measured by Salis *et al.* in (001)-grown QW [38], we get excellent fits of the experimental dependence of the spin relaxation time with $B$ with the single adjustable parameter $\tau_p^* = 0.25$ ps (figure 4b and 5).

In asymmetric (110)-grown QWs, as discussed above, the spin relaxation is also dominated by the DP mechanism. However, the BIA and SIA effective magnetic fields are now orthogonal to each other, see Fig. 6b. The total effective field lies in plane $(x\tilde{z})$ which is obtained from the $(xz)$ plane by rotation around the $x$ axis with the angle $\theta = \arctan[2r_{41}E/(\gamma<k_z^2>)]$. As a result, the growth direction $z$ is not an eigen axis of the spin-relaxation-rate tensor as well as the $g$-factor tensor. This leads to a dynamical coupling of the in-plane and out-of-plane components of the electron spin and yields an unusual dependence of the spin relaxation time on the longitudinal magnetic field $B\|z$ [31,39]. In a simple picture, the mechanism is the following. If $B=0$, the photogenerated electron spin, initially oriented along $z$ (the growth axis), can be decomposed into its projections $S_{\tilde{z}}$ and $S_{\tilde{y}}$ along the eigen axes $\tilde{z}$ (lying in the plane of vectors $\Omega_k$) and $\tilde{y}$ (normal to the plane of vectors $\Omega_k$) of the spin-relaxation-rate tensor. The components $S_{\tilde{z}}$ and $S_{\tilde{y}}$ decay independently and at different rates. $S_{\tilde{y}}$ decays rapidly whereas $S_{\tilde{z}}$ keeps its orientation for a much longer time since the corresponding DP mechanism is suppressed and determined by SIA only. When an external magnetic field $B$ is applied along the $z$ axis, the Larmor precession deflects the electron spin from the long-lifetime axis $\tilde{z}$ and speeds up the DP spin relaxation yielding a reduction of the average spin. As a result, one can expect a *decrease* of the electron spin relaxation time with the magnetic field. For larger magnetic fields, the cyclotron motion slows down the DP electron spin relaxation. Taking into account both effects, the variation of the electron spin relaxation with the magnetic field can exhibit very unusual behavior depending on the ratio of the BIA and SIA terms, cyclotron and Larmor frequencies.

Because of the presence of the internal and external magnetic fields, the electron spin decay is not mono-exponential. This behavior is observed in experiment, see Fig. 3a and the figure inset. In order to quantitatively compare the experimental results with the above microscopic theory, we plot in Fig. 5 the experimental dependence of the effective spin lifetime $T_{zz}^{\text{exp}}$ extracted from the TRKR kinetics as a function of the magnetic field. The effective spin lifetime is defined as $T_{zz}^{\text{exp}} \equiv \int_0^\infty \theta_{Kerr}(t)/\theta_{Kerr}(0) dt$ where $\theta_{Kerr}$ is the Kerr rotation signal. This dependence can be compared to the calculated one:

$$S_z(B_z)/G_z = T_{zz}(B_z), \qquad (10)$$

where $G_z$ is the spin generation rate and $T_{zz}(B_z)$ is the component of the spin lifetime tensor. The time $T_{zz}$ can be obtained by solving the stationary form of Eq. (4). Such a calculation yields

$$T_{zz}(B_z) = \frac{\left(\Gamma_{xx}\Gamma_{yy} + \Omega'^2_{L,z}\right)}{\Gamma_{xx}(\Gamma_{yy}\Gamma_{zz} - \Gamma_{yz}^2) + (\Gamma_{yy}\Omega'^2_{L,y} + \Gamma_{zz}\Omega'^2_{L,z} + 2\Gamma_{yz}\Omega'_{L,y}\Omega'_{L,z})}, \qquad (11)$$

where the components of the spin-relaxation-rate tensor have the form

$$\Gamma_{xx} = \Gamma_{yy} = \frac{m^*(\gamma^2 <k_z^2>^2 + 4r_{41}^2 E^2)\tau_p^*}{\hbar^4(1+\omega_c^2\tau_p^{*2})} k_B T, \qquad (12)$$

$$\Gamma_{yz} = \frac{2m^*\gamma <k_z^2> r_{41} E \tau_p^*}{\hbar^4(1+\omega_c^2\tau_p^{*2})} k_B T, \qquad \Gamma_{zz} = \frac{8m^* r_{41}^2 E^2 \tau_p^*}{\hbar^4(1+\omega_c^2\tau_p^{*2})} k_B T,$$

and the Larmor frequency components are given by

$$\Omega_{L,y} = \frac{g_{yz}\mu_0 B_z}{\hbar} + \Gamma_{yz}\omega_c\tau_p^*, \qquad \Omega_{L,z} = \frac{g_{zz}\mu_0 B_z}{\hbar} + \frac{\Gamma_{zz}\omega_c\tau_p^*}{2}. \qquad (13)$$

We note that, in QWs with $C_s$ symmetry, the electron $g$ factor is strongly anisotropic: the components $g_{xx}$, $g_{yy}$, $g_{zz}$, $g_{yz}$ and $g_{zy}$ can be nonzero.

The full line in Fig. 5 displays the fit of the experimental data by Eq. (11). We obtain a very good fit of the magnetic field dependence $T_{zz}(B_z)$ using the Rashba parameter $\alpha = 3.1$ meV·Å, the electron momentum relaxation time $\tau_p^* = 0.33$ ps and the off-diagonal $g$-factor $g_{yz} = 0.06$. The Dresselhaus coefficient $\gamma$ and the diagonal $g$-factor component $g_{zz}$ are non-adjustable parameters. We use the same Dresselhaus coefficient as that used in Fig. 4 for the (001) QW: $\gamma = 11$ eV·Å$^3$ [38] and the $g_{zz} = -0.2$ as deduced from experimental and theoretical studies in (001) QWs [40-42]. Note that $\alpha = 3.1$ meV·Å corresponds to the electric field $E \sim 60$ kV/cm using the Rashba coefficient $r_{41} = 6$ e·Å$^2$ calculated by tight-binding method for 8 nm-width QW [43]. This electric field exactly coincides with the field $E = 2\pi e N_d/\varepsilon$ produced by the layer of charged impurities of the density $N_d = 8 \times 10^{11}$ cm$^{-2}$, where $\varepsilon \approx 13$ is the dielectric constant of GaAs. Using the same parameters we have calculated in figure 3b the electron spin dynamics $S_z(t)$ for different magnetic field values. As expected and in agreement with the experimental results we observe non-monoexponential decay times and a crossing point of the different curves for small magnetic field values (compare the inset of figure 3a and figure 3b). In the experiments the oscillations are less pronounced compared to the calculated ones probably due to inhomogeneous contributions in the effective magnetic fields and *g* factors [44].

In conclusion we have measured and calculated the magnetic field dependence of the electron spin dynamics in both symmetric and asymmetric GaAs/AlGaAs quantum wells grown on (110) substrates. In the latter structure, characterized by the point group $C_s$, the growth direction is not the principle axis of the spin relaxation tensor. The measured dependence of the spin relaxation on a magnetic field parallel to the growth axis is well described by the model based on the interplay of the Dresselhaus and Rashba effective magnetic fields which are perpendicular to each other for this quantum well crystalline orientation, leading to a non exponential spin decay.

**Acknowledgments**
Part of this work was supported by the France-China NSFC-ANR research project SPINMAN (Grant No. 10911130356), CAS Grant No. 2011T1J37, National Science Foundation of China (Grant No. 11174338) and National Basic Research Program of China (2009CB930502), Russian Foundation for Basic Research, RF President Grants No. MD-2062.2012.2 and No. NSh-5442.2012.2, EU projects POLAPHEN and SPANGL4Q, and the Foundation "Dynasty". The LIA ILNACS is also gratefully acknowledged.

**Figure Captions:**

Figure 1: Kerr rotation dynamics at $T$ = 150 K in the symmetrical (Sym) and asymmetrical (Asym) GaAs/AlGaAs quantum wells grown on (110) substrates. The pump laser pulse power is $P$ = 5 mW. The insets represent the potential profile in the QW region.

Figure 2: Excitation power dependence of the electron spin lifetime in: (a) the asymmetrical and (b) symmetrical quantum well grown on (110) substrate measured at $T$=150 K.

Figure 3: (a) Kerr rotation dynamics in the (110) asymmetrical quantum well for magnetic fields in the range 0 – 1 T. Inset zooms the spin dynamics in the temporal range $t$ = 250 – 500 ps without and with a 0.3 Tesla applied magnetic field. (b) Time evolution of the electron spin density calculated after Eq. (4), parameters used are given in the text.

Figure 4: (a) Kerr rotation dynamics measured in the (001) symmetrical quantum well for magnetic fields in the range 0 – 1 T. (b) Time evolution of the electron spin density calculated after Eq. (4), parameters used are given in the text.

Figure 5: Effective spin lifetimes measured (symbols) and calculated (solid lines) as a function of the magnetic field. Black squares: asymmetrical (110) quantum well; the Rashba and Dresselhaus contributions are orthogonal. White circles: symmetrical (001) quantum well; the Rashba and Dresselhaus contributions both lie in the quantum well plane. Solid lines: model for both cases, calculated with equations (11) and (8,9) respectively.

Figure 6: The dependence of the effective magnetic field on the wave vector in (a) symmetric (001)-grown QWs and (b) asymmetric (110)-grown QWs. In (001) QWs, the effective field lies in the QW plane, the corresponding principal axes of spin relaxation tensor are $x \parallel [1,\bar{1},0]$, $y \parallel [1,1,0]$, and $z \parallel [0,0,1]$. In (110) QWs, the effective field lies in the plane ($x\tilde{z}$), the corresponding principal axes of spin relaxation tensor are $x$, $\tilde{y}$ and $\tilde{z}$.

Figure 1

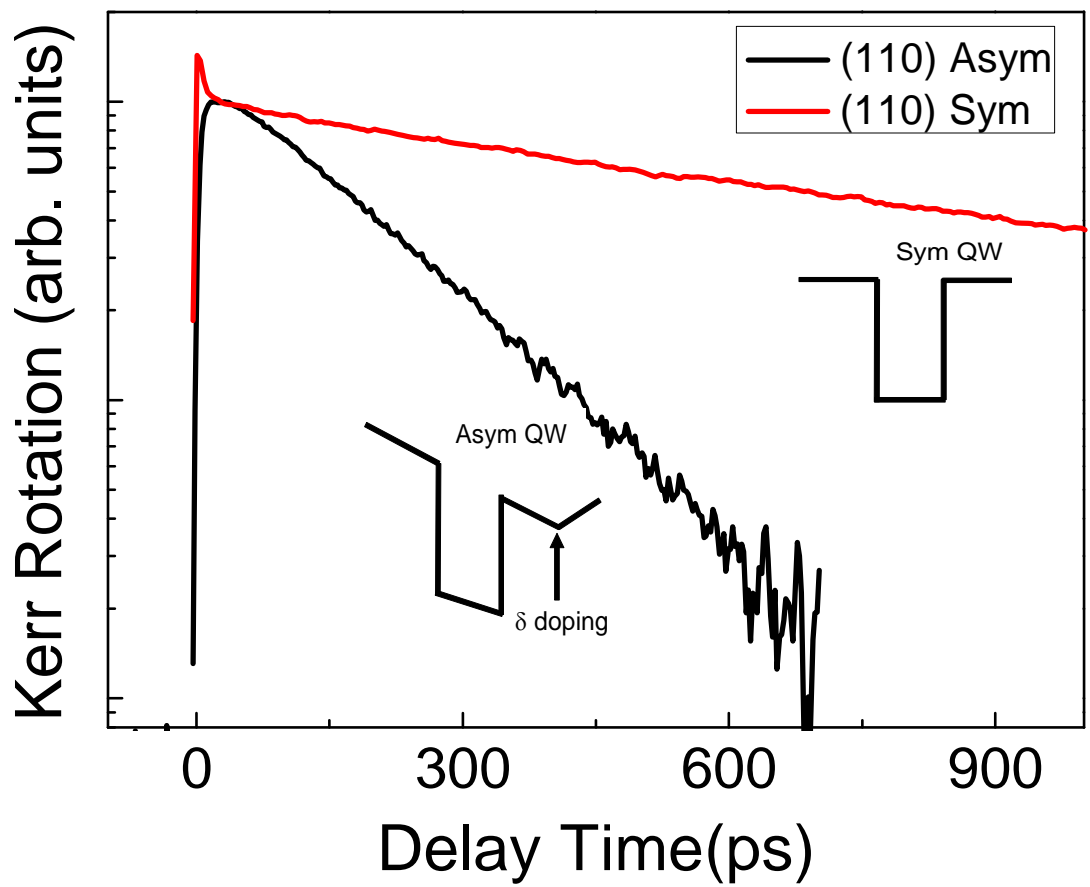

Figure 2a

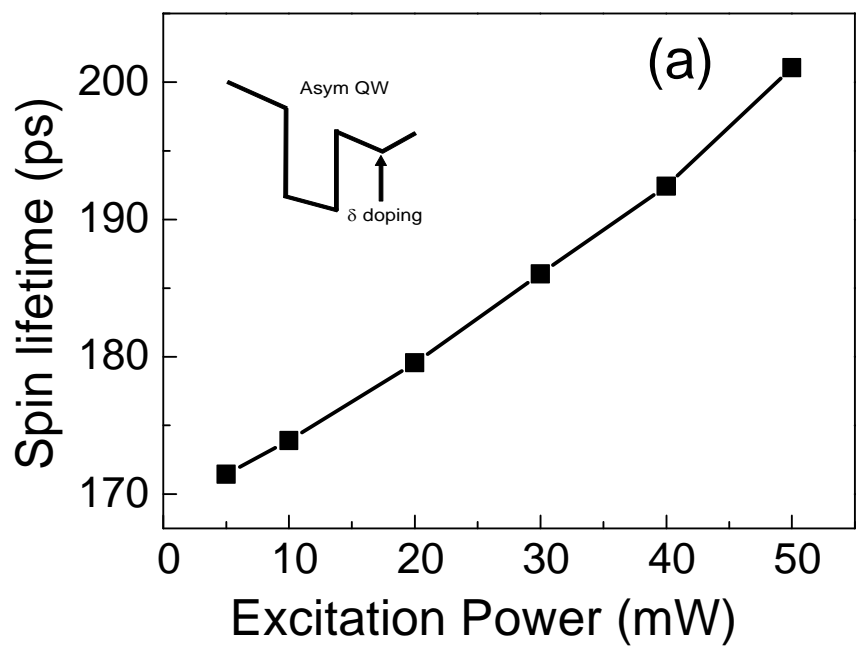

Figure 2b

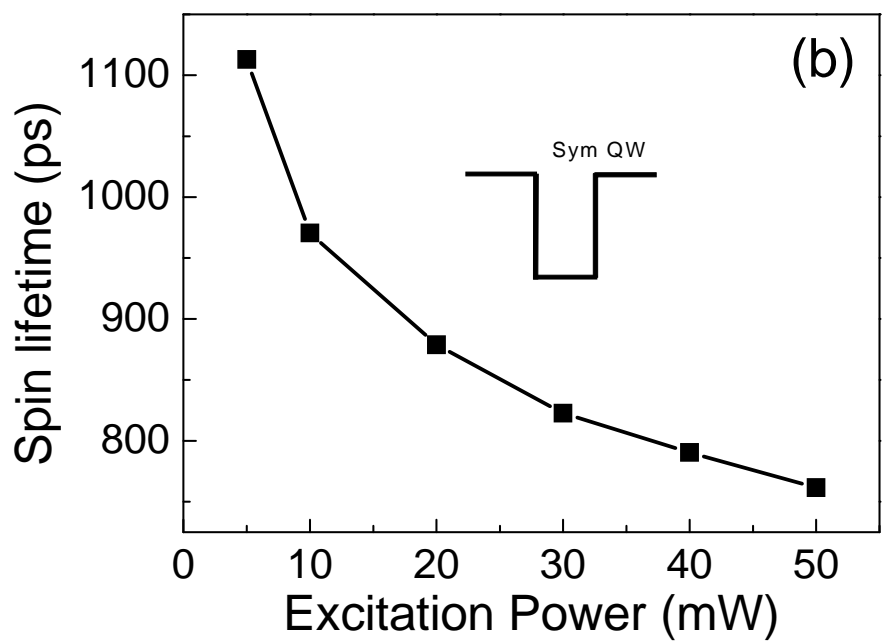

Figure 3a

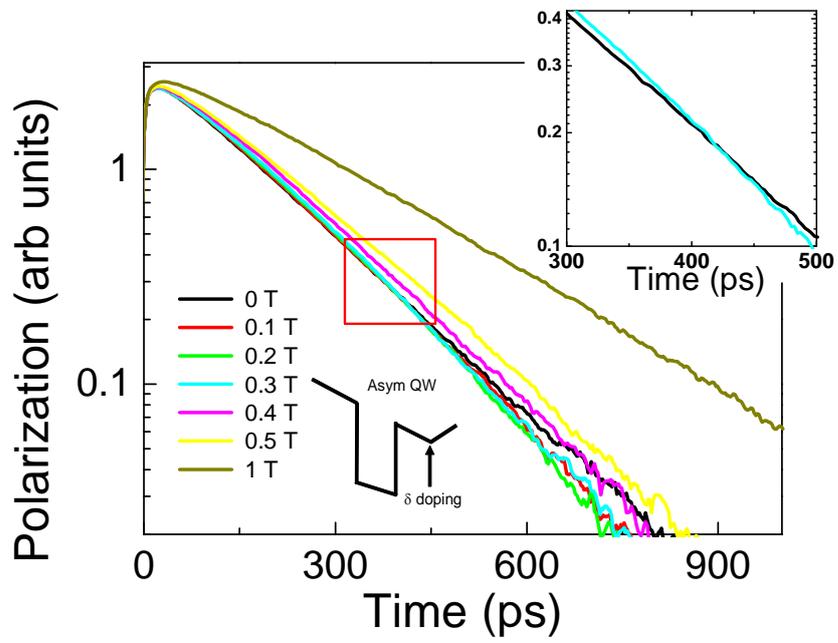

Figure 3b

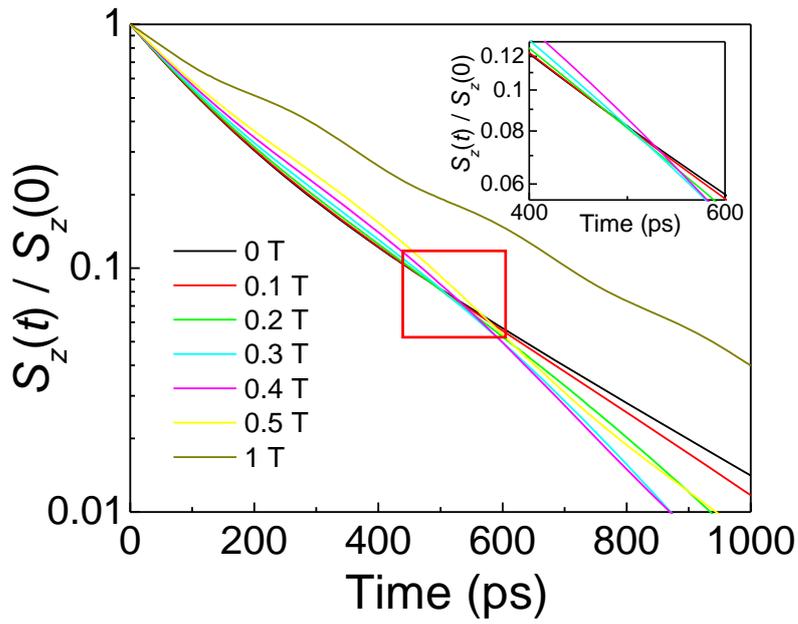

Figure 4a

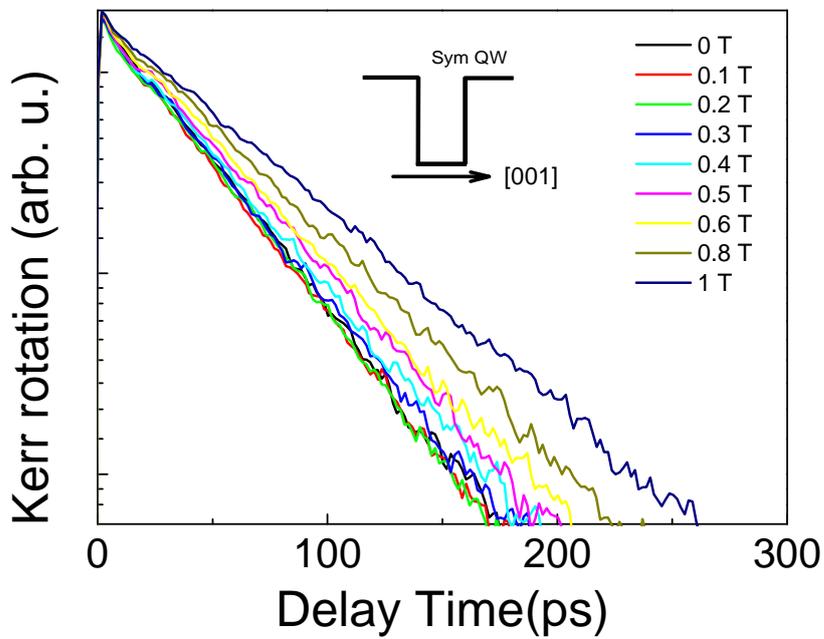

Figure 4b

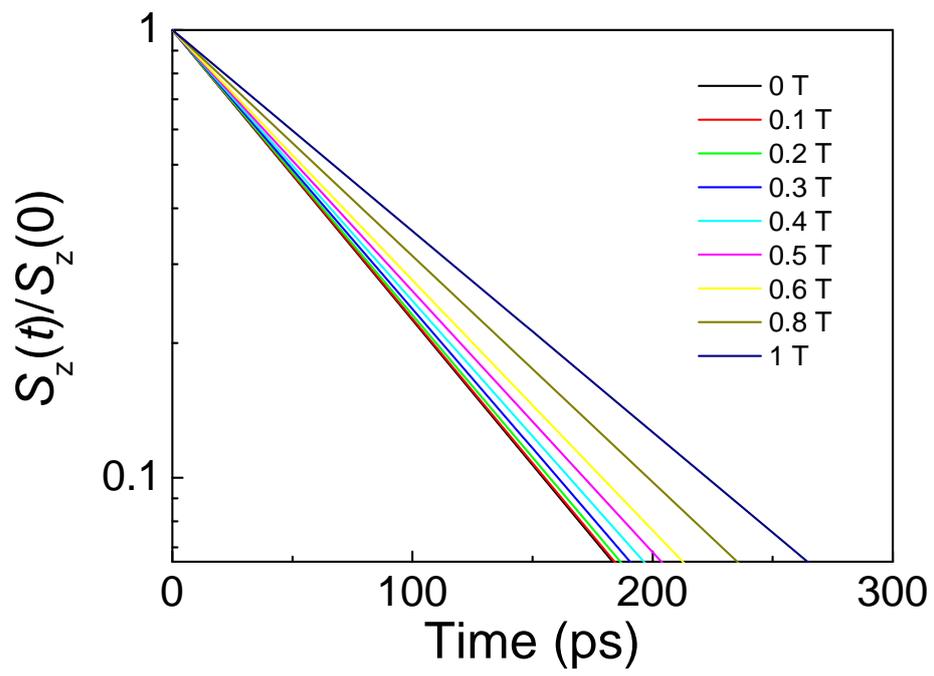

Figure 5:

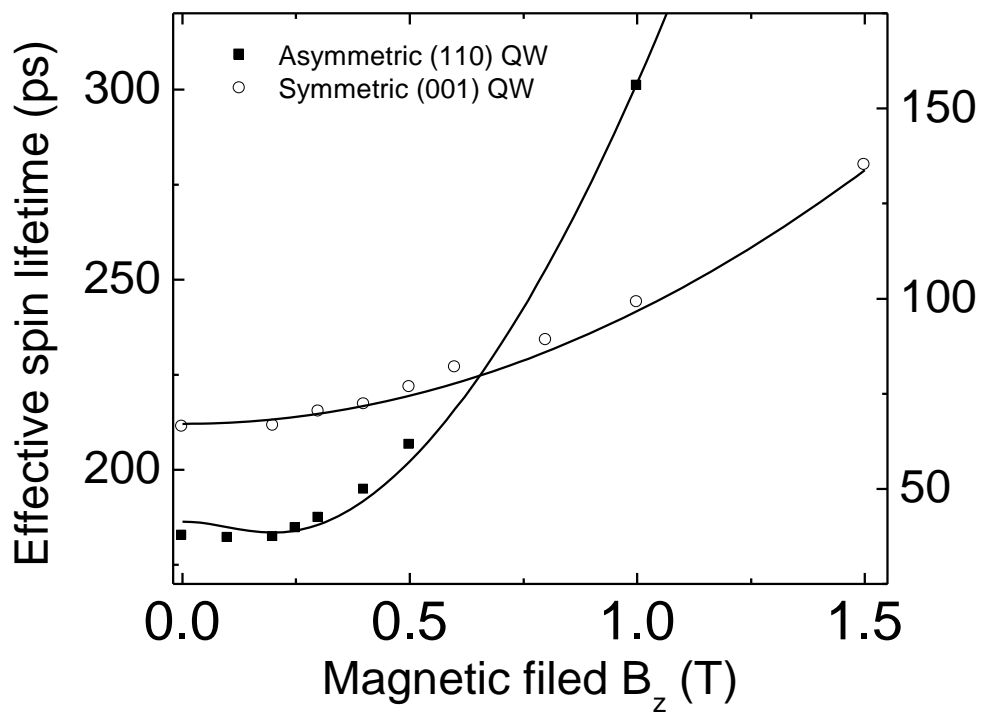

Figure 6:

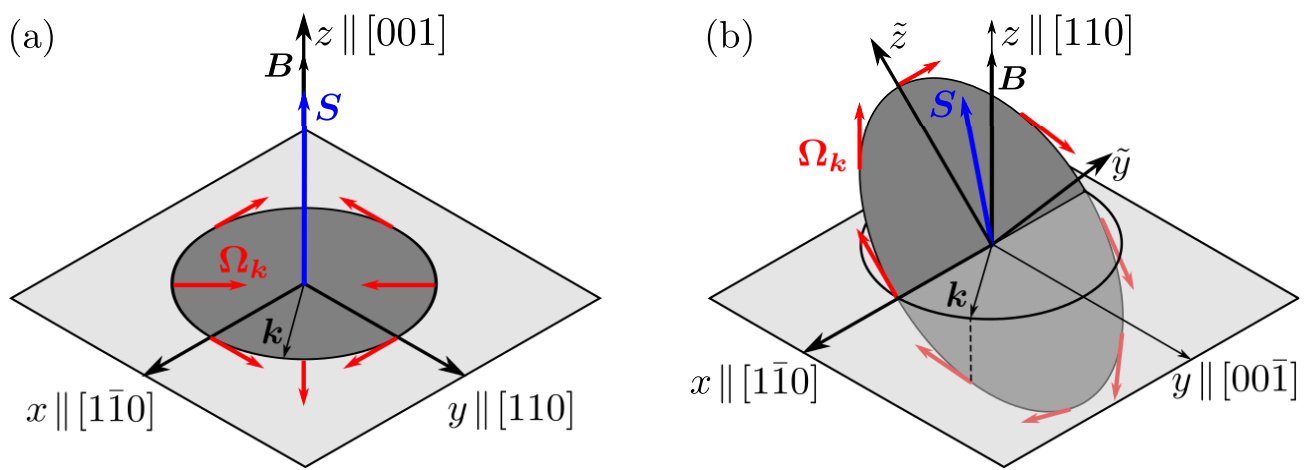